\documentclass[aps,prb,showpacs,twocolumn]{revtex4}
\usepackage{amsmath,amssymb,mathrsfs,latexsym,graphicx,psfrag}
\usepackage[varg]{txfonts}

\newcommand{\bs}[1]{\boldsymbol{#1}}
\newcommand{\braket}[2]{\left\langle #1 | #2 \right\rangle}
\newcommand{\Bigbraket}[2]{\Big\langle #1 \Big| #2 \Big\rangle}

\newcommand{\bra}[1]{\left\langle#1\right|}
\newcommand{\ket}[1]{\left|#1\right\rangle}

\newcommand{\comm}[2]{\left[#1,#2\right]}

\newcommand{\abs}[1]{\left|#1\right|}

\newcommand{\parder}[2]{\frac{\partial #1}{\partial #2}}

\renewcommand{\strut}{\rule[-.9\baselineskip]{0pt}{2.5\baselineskip}}
\def\s{\scriptscriptstyle}

\def\ie{{\it i.e.},\ }

\allowdisplaybreaks[1]
\begin{document}
\title{Landau Level Quantization on the Sphere} \author{Martin
  Greiter} \affiliation{Institut f\"ur Theorie der Kondensierten
  Materie and DFG Center for Functional Nanostructures (CFN), KIT,
  Campus S\"ud, D 76128 Karlsruhe} \pagestyle{plain} \date{\today}

\begin{abstract}
  It is well established that the Hilbert space for charged particles
  in a plane subject to a uniform magnetic field can be described by
  two mutually commuting ladder algebras.  We propose a similar
  formalism for Landau level quantization on a sphere 
  involving two mutually commuting SU(2) algebras.
\end{abstract}

\maketitle

\section{Introduction}
The formalism for Landau level quantization in a spherical geometry,
\ie for the dynamics of a charged particle on the surface of a sphere
with radius $R$, in a magnetic (monopole) field, was pioneered by
Haldane~\cite{haldane83prl605} in 1983 as an alternative geometry for
the formulation of fractionally quantized Hall
states~\cite{ChakrabortyPietilainen95}.  In comparison to the disk
geometry used by Laughlin~\cite{laughlin83prl1395} when he originally
proposed the Jastrow-type wave functions for the ground state, the
sphere has the advantage that it does not have a boundary.  At the
same time, it does not display the topological degeneracies associated
with the torus geometry (\ie a plane with periodic boundary
conditions), which has genus one~\cite{haldane-85prb2529}.  These two
properties make the spherical geometry particularly suited for
numerical work on bulk properties of quantized Hall states.

Haldane~\cite{haldane83prl605}, however, worked out the formalism only
for the lowest Landau level, and never generalized it to higher Landau
levels, even though these became more and more important as time
passed.  The presently most vividly discussed quantum Hall state, the
Pfaffian
state\cite{moore-91npb362,greiter-91prl3205,
  moeller08prb075319,baraban-09prl076801,bishara-09prb155303,moore09p82,stern10n187,storni-10prl076803,wojs-10prl086801,thomale-10prl180502,wojs-10prl105.096802}
at Landau level filling fraction $\nu=5/2$, is observed only in the
second Landau level.

In this article, we first review Haldane's formalism for the lowest
Landau level~\cite{haldane83prl605,fano-86prb2670}, and then
generalize it to the full Hilbert space, which includes higher Landau
levels as well.  The key insight permitting this generalization is
that there is not one, but that there are two mutually commuting SU(2)
algebras with spin $s$, one for the cyclotron variables and one for
the guiding center variables.  The formalism we develop will prove
useful for numerical studies of fractionally quantized Hall states
involving higher Landau levels.  In particular, it will instruct us
how to calculate pseudopotentials~\cite{haldane83prl605} for higher
Landau levels on the sphere, which we will discuss as well.  Finally,
we will present a convenient way to write the wave function for
$M$ filled Landau levels on the sphere.

\section{Haldane's formalism}
Following Haldane~\cite{haldane83prl605}, we assume a radial magnetic
field of strength
\begin{equation}
  \label{eq:b}
  B=\frac{\hbar c s_0}{eR^2}\qquad(e>0).
\end{equation}
The number of magnetic Dirac flux quanta through the surface of the sphere is
\begin{equation}
  \label{eq:phitot}
  \frac{\Phi_\text{tot}}{\Phi_0}=\frac{4\pi R^2B}{{2\pi\hbar c}/{e}} 
  =2s_0,
\end{equation}
which must be integer due to Dirac's monopole quantization 
condition~\cite{dirac31prsla60}.  In the following, we take $\hbar=c=1$.

The Hamiltonian is given by 
\begin{equation}
  \label{eq:ham1}
  H\;=\;\frac{\bs{\Lambda}^2}{2MR^2}
  \;=\;\frac{\omega_\text{c}}{2 s_0}\bs{\Lambda}^2,
\end{equation}
where $\omega_\text{c}=eB/M$ is the cyclotron frequency,
\begin{equation}
  \label{eq:lambda1}
  \bs{\Lambda}\;=\;\bs{r}\times(-i\nabla + e\bs{A}(\bs{r}))
\end{equation}
is the dynamical angular momentum, $\bs{r}=R\bs{e}_r$, and
$\nabla\times\bs{A}=B\bs{e}_r$.  With
\eqref{eq:app-sprhs}--\eqref{eq:app-spnabla} from the appendix
we obtain
\begin{equation}
  \label{eq:lambda2}
  \bs{\Lambda}\;=\;-i\left(
    \bs{e}_\varphi\parder{}{\theta}
    -\bs{e}_\theta\frac{1}{\sin\theta}\parder{}{\varphi}
  \right) + eR\,(\bs{e}_r\times\bs{A}(\bs{r})).
\end{equation}
Note that 
\begin{equation}
  \label{eq:edotlambda}
  \bs{e}_r\bs{\Lambda}=\bs{\Lambda}\bs{e}_r=0,
\end{equation}
as one can easily verify with \eqref{eq:app-spderivatives}.
%
%
The commutators of the Cartesian components of $\bs{\Lambda}$ with themselves
and with $\bs{e}_r$ can easily be evaluated using \eqref{eq:lambda2} and 
\eqref{eq:app-spe}--\eqref{eq:app-spderivatives}.  This yields
\begin{gather}
  \label{eq:lambdacomm}
  \comm{\Lambda^i}{\Lambda^j}
  =i\epsilon^{ijk}(\Lambda^k- s_0\,e_r^k), 
  \\[5pt] \label{eq:lambdaecomm}
  \comm{\Lambda^i}{e_r^j}=i\epsilon^{ijk}e_r^k,
\end{gather}
where $i,j,k=x,y,\ \text{or}\ z$, and $e_r^k$ is
the $k$-th Cartesian coordinate of $\bs{e}_r$.  From
\eqref{eq:edotlambda}--\eqref{eq:lambdaecomm},
we see that that the operator
\begin{equation}
  \label{eq:L1}
  \bs{L}=\bs{\Lambda}+ s_0\bs{e}_r
\end{equation}
is the generator of rotations around the origin,
\begin{equation}
  \label{eq:lambdaplusecomm}
  \comm{L^i}{X^j}=i\epsilon^{ijk}X^k
\quad\text{with}\quad \bs{X}=\bs{\Lambda},\;\bs{e}_r,\;\text{or}\;\bs{L},
\end{equation}
and hence the angular momentum.  As it satisfies the angular momentum
algebra, it can be quantized accordingly.  Note that $\bs{L}$ has a
component in the $\bs{e}_r$ direction:
\begin{equation}
  \label{eq:Ldoter}
  \bs{L}\bs{e}_r=\bs{e}_r\bs{L}= s_0.
\end{equation}
If we take the eigenvalue of $\bs{L}^2$ to be $s(s+1)$, this implies
$s=s_0+n$, where $n=0,1,2,\ldots$ is a non-negative integer (while $s$
and $s_0$ can be integer or half integer, according to number of Dirac
flux quanta through the sphere).

With \eqref{eq:L1} and \eqref{eq:edotlambda}, we obtain
\begin{equation}
  \label{eq:lambda^2L^2}
  \bs{\Lambda}^2=\bs{L}^2-s_0^2.
\end{equation}
The energy eigenvalues of \eqref{eq:ham1} are hence
\begin{eqnarray}
  \label{eq:e}
  E_n\!&=&\!\frac{ \omega_\text{c}}{2s_0}
   \left[s(s+1) - s_0^2\right]\nonumber\\[0.3\baselineskip]
   \!&=&\!\frac{ \omega_\text{c}}{2s_0}
   \left[(2n+1)s_0 + n(n+1) \right]\nonumber\\[0.3\baselineskip]
   \!&=&\! \omega_\text{c}
   \left[\left(n+\frac{1}{2}\right) + \frac{n(n+1)}{2s_0} \right].
\end{eqnarray}
The index $n$ hence labels the Landau levels.

To obtain the eigenstates of \eqref{eq:ham1}, we have to choose a
gauge and then explicitly solve the eigenvalue equation.  We choose the
latitudinal gauge
\begin{equation}
  \label{eq:gauge}
  \bs{A}=-\bs{e}_\varphi\frac{s_0}{eR}\cot\theta.
\end{equation}
The singularities of $\bs{B}=\nabla\times\bs{A}$ at the poles are
without physical significance.  They describe infinitly thin solenoids
admitting flux $s_0\Phi_0$ each and reflect our inability to formulate
a true magnetic monopole.

The dynamical angular momentum \eqref{eq:lambda2} becomes
\begin{equation}
  \label{eq:lambda3}
  \bs{\Lambda}\;=\;-i\left[
    \bs{e}_\varphi\parder{}{\theta}
    -\bs{e}_\theta\frac{1}{\sin\theta}
    \left(\parder{}{\varphi}-is_0\cos\theta\right)
  \right].
\end{equation}
With \eqref{eq:app-spderivatives} we obtain
\begin{equation}
  \label{eq:lambda^2}
  \bs{\Lambda}^2\;=\;
     -\frac{1}{\sin\theta}\parder{}{\theta}
     \left(\sin\theta\parder{}{\theta}\right)
     -\frac{1}{\sin^2\theta}
     \left(\parder{}{\varphi}-is_0\cos\theta\right)^2.
\end{equation}

To formulate the eigenstates, Haldane~\cite{haldane83prl605}
introduced spinor coordinates for the particle position,
\begin{equation}
  \label{eq:uv}
  u=\cos\frac{\theta}{2} \exp\left({\frac{i\varphi}{2}}\right),\quad
  v=\sin\frac{\theta}{2} \exp\left({-\frac{i\varphi}{2}}\right),
\end{equation}
such that
\begin{equation}
  \label{eq:Omegauv}
  \bs{e}_r
  \;=\;\bs{\Omega}(u,v)\;\equiv\;(u,v)\,\bs{\sigma}\!\left(\!\!
    \begin{array}{c}
      \bar u\\[2pt]\bar v
    \end{array}\!\!\right)\!,
\end{equation}
where $\bs{\sigma}=(\sigma_x,\sigma_y,\sigma_z)$ is the vector consisting of
the three Pauli matrices
\begin{equation}
  \label{eq:Pauli}
  \sigma_x=\!\left(\!\!
    \begin{array}{cc}
      0&1\\1&0 
    \end{array}\!\!\right),\quad
  \sigma_y=\!\left(\!\!
    \begin{array}{cc}
      0&-i\\i&0 
    \end{array}\!\!\right),\quad
  \sigma_z=\!\left(\!\!
    \begin{array}{cc}
      1&0\\0&-1 
    \end{array}\!\!\right).
\end{equation}

In terms of these, a complete, orthogonal
basis of the states spanning the lowest Landau
level ($n=0$, $s=s_0$) is given by
\begin{equation}
  \label{eq:LLLbasis}
  \psi_{m,0}^{s}(u,v) = 
  u^{s+m}v^{s-m},
\end{equation}
with $m=-s,s+1,\ldots,s.$  For these states,
\begin{eqnarray}
  \label{eq:psi0eq}
  L^z\psi_{m,0}^{s}&\!=\!&m\,\psi_{m,0}^{s},
  \nonumber\\ [0.2\baselineskip]
  H\psi_{m,0}^{s}&\!=\!&\frac{1}{2}\omega_\text{c}\,\psi_{m,0}^{s}.
\end{eqnarray}
To verify \eqref{eq:psi0eq}, we consider the action of 
\eqref{eq:lambda^2} on the more general basis states
\begin{eqnarray}
  \label{eq:mpbasis}
  \phi_{m,p}^{s}(u,v)
  &\!=\!&\left(\cos\frac{\theta}{2}\right)^{s+m} 
         \left(\sin\frac{\theta}{2}\right)^{s-m} 
         e^{i(m-p)\varphi}
  \nonumber\\ [0.5\baselineskip]
  &\!=\!& \left\{\begin{array}{ll}
    {\bar v}^{-p}\,u^{s+m}\,v^{s-m+p},\quad & \text{for}\ p<0,\\[5pt] 
    {\bar u}^{p}\,u^{s+m-p}\,v^{s-m}, \quad & \text{for}\ p\ge 0.
    \end{array}\right.
\end{eqnarray}
This yields
\begin{eqnarray}
  \label{eq:Lambda^2mpbasis}
    &&\hspace{-14pt}\bs{\Lambda}^2\,\phi_{m,p}^{s}
  \nonumber\\[0.3\baselineskip]
  \!&\!=\!&\!\!\left[s
    -\left(\frac{s\cos\theta-m}{\sin\theta}\right)^2
    +\left(\frac{s_0\cos\theta-m+p}{\sin\theta}\right)^2
  \right]\phi_{m,p}^{s}
  \nonumber\\[0.3\baselineskip]
   \!&\!=\!&\!\left[s
     +\frac{2(s\cos\theta\!-\!m\!+\!p)(p\!-\!n\cos\theta)
       -(p^2\!-\!n^2\cos^2\theta)}{\sin^2\theta}
   \right] \phi_{m,p}^{s}.
   \nonumber\\
\end{eqnarray}
For $p=n=0$, this clearly reduces to
$\bs{\Lambda}^2\,\psi_{m,0}^{s}=s\,\psi_{m,0}^{s}$, and hence
\eqref{eq:psi0eq}.
The normalization of \eqref{eq:LLLbasis} can easily be obtained
with the integral
\begin{equation}
  \label{eq:psinorm}
  \frac{1}{4\pi}\int \text{d}\Omega\, \bar u^{s'+m'}\bar v^{s'-m'} u^{s+m}v^{s-m}
  =\frac{(s+m)!\, (s-m)!}{(2s+1)!}\,\delta_{mm'}\delta_{ss'}
\end{equation}
where $\text{d}\Omega=\sin\theta\, \text{d}\theta\, \text{d}\phi$.

To describe particles in the lowest Landau level which are localized at a
point $\bs{\Omega}(\alpha,\beta)$ with spinor coordinates
$(\alpha,\beta)$,
\begin{equation}
  \label{eq:Omegaalphabeta}
  \bs{\Omega}(\alpha,\beta)=(\alpha,\beta)\,\bs{\sigma}\!\left(\!\!
    \begin{array}{c}
      \bar\alpha\\[2pt]\bar\beta
    \end{array}\!\!\right)\!,
\end{equation}
Haldane~\cite{haldane83prl605} introduced ``coherent states'' defined by
\renewcommand{\strut}{\rule[-.6\baselineskip]{0pt}{2\baselineskip}}
\begin{equation}
  \label{eq:CoherentDef}
  \{\bs{\Omega}(\alpha,\beta)\,\bs{L}\}\,\psi_{(\alpha,\beta),0}^{s}(u,v)
  =s\,\psi_{(\alpha,\beta),0}^{s}(u,v).
\end{equation}
In the lowest Landau level, the angular momentum $\bs{L}$ can be written
\begin{equation}
  \label{eq:lLLL}
  \bs{L}=\frac{1}{2}(u,v)\,\bs{\sigma}\!\left(\!\!
    \begin{array}{c}
      \parder{}{u}\\[4pt] \parder{}{v}
    \end{array}\!\!\right)\!.
\end{equation}
Note that $u,v$ may be viewed as Schwinger boson creation, and
$\parder{}{u},\parder{}{v}$ the corresponding annihilation
operators~\cite{schwinger65proc}.
%
The solutions of \eqref{eq:CoherentDef} are given by
\begin{equation}
  \label{eq:Coherent}
  \psi_{(\alpha,\beta),0}^{s}(u,v) = (\bar\alpha u + \bar\beta v)^{2s},
\end{equation}
as one can verify easily with the identity
\begin{equation}
  \label{eq:Biedenharn}
   (\underline{a}\,\bs{\sigma}\,\underline{b})
   (\underline{c}\,\bs{\sigma}\,\underline{d})
      =2 (\underline{a}\,\underline{d})(\underline{c}\,\underline{b})
      -(\underline{a}\,\underline{b})(\underline{c}\,\underline{d}).
\end{equation}
where $\underline{a}$, $\underline{b}$, $\underline{c}$,
$\underline{d}$ are two-component spinors.


\section{Generalization to higher Landau levels}
We will first present the formalism we developed and then motivate it.
In analogy to the two mutually commuting ladder algebras $a,a^\dagger$
and $b,b^\dagger$ in the
plane~\cite{note,landau30zp629,macdonald84prb3550,girvin-83prb4506,Arovas86},
we describe the Hilbert space of a charged particle on a sphere with a
magnetic monopole in the center by two mutually commuting SU(2)
angular momentum algebras.  The first algebra for the cyclotron
momentum $\bs{S}$, consists of operators which allow us raise or lower
eigenstates from one Landau to the next (as $a,a^\dagger$ do in the
plane). The second algebra for the guiding center momentum $\bs{L}$,
consists of operators which rotate the eigenstates on the sphere while
preserving the Landau level index (as $b,b^\dagger$ do in the plane).

The reason this structure has not been discovered long ago may be that
it is possible to obtain the spectrum without introducing $\bs{S}$, as
the eigenvalue of both $\bs{S}^2$ and $\bs{L}^2$ is $s(s+1)$.  The
necessity to introduce $\bs{S}$ is hence not obvious.

We have already seen above that the spinor coordinates $u,v$ and the
derivatives $\parder{}{u},\parder{}{v}$ may be viewed as Schwinger
boson creation and annihilation operators, respectively.  A complete
basis for the eigenstates of $H$ in the lowest Landau level is given
by $u^{m+s}v^{m-s}$, \ie could be expressed in terms of $u$ and
$v$.  For higher Landau levels, the analogy to the plane suggests that
we will need $\bar u,\bar v$ as well.  With the derivatives
$\parder{}{\bar u},\parder{}{\bar v}$, we have a total of four
Schwinger boson creation and annihilation operators.  This suggests
that we span two mutually commuting SU(2) algebras with them.

We will motivate below that the appropriate combinations are
\renewcommand{\strut}{\rule{0pt}{22pt}}
\begin{equation}
\begin{array}{r@{\hspace{5pt}}c@{\hspace{5pt}}l@{\hspace{5pt}}c@{\hspace{5pt}}l}
  S^x + iS^y &=& S^+ &=&\displaystyle
  u\parder{}{\bar v}- v\parder{}{\bar u},\\\strut
  S^x - iS^y &=& S^- &=&\displaystyle
  \bar v\parder{}{u}- \bar u\parder{}{v},\\\strut 
  && S^z  &=&\displaystyle
  \frac{1}{2} \left(
    u\parder{}{u} + v\parder{}{v} 
    - \bar u\parder{}{\bar u}- \bar v\parder{}{\bar v}
  \right),
  \label{eq:Sdef}
\end{array}
\end{equation}
for the cyclotron momentum and
\renewcommand{\strut}{\rule{0pt}{22pt}}
\begin{equation}
\begin{array}{r@{\hspace{5pt}}c@{\hspace{5pt}}l@{\hspace{5pt}}c@{\hspace{5pt}}l}
  L^x + iL^y &=& L^+ &=&\displaystyle
  u\parder{}{v}- \bar v\parder{}{\bar u},\\\strut
  L^x - iL^y &=& L^- &=&\displaystyle
  v\parder{}{u}- \bar u\parder{}{\bar v},\\\strut 
  && L^z  &=&\displaystyle
  \frac{1}{2} \left(
    u\parder{}{u} - v\parder{}{v} 
    - \bar u\parder{}{\bar u}+ \bar v\parder{}{\bar v}
  \right),
  \label{eq:Ldef}
\end{array}
\end{equation}
for the guiding center momentum.  We can write these more compactly as
\begin{eqnarray}
  \label{eq:Spauli}
  \bs{S}&\!=\!&\frac{1}{2}(u,\bar v)\,\bs{\sigma}\!\left(\!\!
    \begin{array}{c}
      \parder{}{u}\\[4pt] \parder{}{\bar v}
    \end{array}\!\!\right)
  -\frac{1}{2}(\bar u,v)\,\bs{\sigma}^\text{T}\!\left(\!\!
    \begin{array}{c}
      \parder{}{\bar u}\\[4pt] \parder{}{v}
    \end{array}\!\!\right),\\[0.8\baselineskip]
  \label{eq:Lpauli}
  \bs{L}&\!=\!&\frac{1}{2}(u,v)\,\bs{\sigma}\!\left(\!\!
    \begin{array}{c}
      \parder{}{u}\\[4pt] \parder{}{v}
    \end{array}\!\!\right)
  -\frac{1}{2}(\bar u,\bar v)\,\bs{\sigma}^\text{T}\!\left(\!\!
    \begin{array}{c}
      \parder{}{\bar u}\\[4pt] \parder{}{\bar v}
    \end{array}\!\!\right),
\end{eqnarray}
where $\bs{\sigma}^\text{T}=(\sigma_x,-\sigma_y,\sigma_z)$ is the vector
consisting of the three transposed Pauli matrices.

From \eqref{eq:Spauli} and \eqref{eq:Lpauli}, we see that both
$\bs{S}$ and $\bs{L}$ obey the SU(2) angular momentum algebras
\begin{equation}
  \label{eq:SLsu2}
  \comm{S^i}{S^j}=i\epsilon^{ijk}S^k,\qquad 
  \comm{L^i}{L^j}=i\epsilon^{ijk}L^k. 
\end{equation}
With \eqref{eq:Sdef} and \eqref{eq:Ldef}, it is easy to show that
the two algebras are mutually commutative,
\begin{equation}
  \label{eq:SLcomm}
  \comm{S^i}{L^j}=0\quad\text{for all}\ i,j.
\end{equation}
For $\bs{S}^2$ and $\bs{L}^2$ we find
\begin{equation}
  \label{eq:s^2}
  \bs{L}^2=\bs{S}^2=S(S+1),
\end{equation}
with
\begin{equation}
  \label{eq:s}
  S=\frac{1}{2} \left(
    u\parder{}{u} + v\parder{}{v} 
    + \bar u\parder{}{\bar u}+ \bar v\parder{}{\bar v}
  \right).
\end{equation}
The component of $\bs{L}$ normal to the surface of the sphere is
\begin{eqnarray}
  \label{eq:OmegaL}
  \bs{e}_r\bs{L}=\bs{\Omega}(u,v)\,\bs{L}=S^z,
\end{eqnarray}
which is easily verified with \eqref{eq:Lpauli},
\eqref{eq:Omegauv}, \eqref{eq:Biedenharn},
%
and
\begin{gather}
   \label{eq:Omegaubarvbar}
   \bs{\Omega}(u,v)\;=\;(\bar u,\bar v)\,\bs{\sigma}^{\text{T}}\!\left(\!\!
    \begin{array}{c}
      u\\[2pt] v
    \end{array}\!\!\right)\!,\\
   (\underline{a}\,\bs{\sigma}^{\text{T}}\,\underline{b})
   (\underline{c}\,\bs{\sigma}^{\text{T}}\,\underline{d})
      =2 (\underline{a}\,\underline{d})(\underline{c}\,\underline{b})
      -(\underline{a}\,\underline{b})(\underline{c}\,\underline{d}).
\end{gather}
It implies that the physical Hilbert space is limited to states 
with $S^z$ eigenvalue $s_0$, 
\renewcommand{\strut}{\rule[-.7\baselineskip]{0pt}{2.1\baselineskip}}
\begin{equation}
  \label{eq:Sz}
  S^z\psi = s_0\psi \quad\text{for all eigenstates}\ \psi .
\end{equation}

With \eqref{eq:s^2}--\eqref{eq:OmegaL}, we write
\begin{eqnarray}
  \label{eq:h2}
  H&\!=\!&\frac{\omega_\text{c}}{2 s_0}\left(
    \bs{L}^2-(\bs{e}_r\bs{L})^2\right)
  \nonumber\\[0.3\baselineskip]
  &\!=\!&\frac{\omega_\text{c}}{2 s_0}\left(\bs{S}^2-{(S^z)}^2\right)
  \nonumber\\[0.3\baselineskip]
  &\!=\!&\frac{\omega_\text{c}}{4 s_0}\left(S^+S^-+S^-S^+\right).
\end{eqnarray}
With 
$\comm{S^+}{S^-}=2 S^z$ 
and \eqref{eq:Sz}, we obtain
\renewcommand{\strut}{\rule[-1.2\baselineskip]{0pt}{2.8\baselineskip}}
\begin{equation} 
   \label{eq:ham3}
    H\;=\;\omega_\text{c}\left(\frac{1}{2 s_0} S^-S^+ + \frac{1}{2}\right).
\end{equation}
This is our main result.
The operators $S^-$ and $S^+$ hence play the role of Landau level
raising and lowering operators, respectively, as $a^\dagger$ and $a$
do in the plane~\cite{Arovas86}.  At the same time,
the raising operator $S^-$ lowers the eigenvalue of $S^z$ (\ie $s_0$)
by one, as
\begin{equation}
  \label{eq:S+S-comm}
  \comm{S^z}{S^-}=-S^z.
\end{equation}
This has to be taken into account when constructing the Hilbert space.

The guiding center momentum $\bs{L}$ generates rotations of the states
within each Landau level around the sphere, while leaving the Landau
level structure unaltered.
%
%
Note that the seemingly unrelated forms \eqref{eq:Lpauli} and
\eqref{eq:L1} of $\bs{L}$ describe the same operator, as both
generate identical rotations around the sphere. 

The basis states \eqref{eq:LLLbasis} are obviously eigenstates of
\eqref{eq:ham3} with energy $\frac{1}{2}\omega_\text{c}$.  To lift
them into the $(n+1)$-th Landau level, we only have to increase the
flux from $s_0$ to $s=s_0+n$, and then apply $(S^-)^n$:
\begin{equation}
  \label{eq:LLbasis}
  \psi_{m,n}^{s}(u,v) = (S^-)^n\, \psi_{m,0}^{s}(u,v).
\end{equation}
where $s=s_0+n$ and $m=-s,\ldots,s$.  The states $\psi_{m,n}^{s}(u,v)$
constitute a complete, orthogonal basis for the $(n+1)$-th Landau
level on a sphere in a monopole field with $2s_0$ Dirac flux quanta
through its surface.  

We will now show that the states \eqref{eq:LLbasis} are indeed
eigenstates of \eqref{eq:ham1} with energy \eqref{eq:e}.  Note
first that since
\begin{equation*}
  \label{eq:psiL0}
  \psi_{m,0}^{s}(u,v)=\frac{(2s-m)!}{(2s)!}\,(L^-)^m\,\psi_{S,0}^{s}(u,v)
\end{equation*}
and $\comm{S^-}{L^-}=\comm{H}{L^-}=0$, it is sufficient to show that
\begin{equation*}
  \label{eq:LLbasis0}
  \psi_{0,n}^{s}(u,v) = (S^-)^n\, 
  \underbrace{\psi_{s,0}^{s}(u,v)}_{\textstyle =u^{2s}} 
  = \frac{(2s)!}{(2s-n)!}\, {\bar v}^n u^{2s-n}
\end{equation*}
is an eigenstate. 
This is (up to a normalization) 
equal to our earlier basis state $\phi_{m,p}^{s}$ (see
\eqref{eq:mpbasis}) with $m=s-n$, $p=-n$.  With
\eqref{eq:Lambda^2mpbasis}, we find
\begin{eqnarray}
  \label{eq:Lambda^2LLbasis0}
  &&\hspace{-17pt}\bs{\Lambda}^2\,\psi_{0,n}^{s}
  \nonumber\\[0.3\baselineskip]
  \!&\!=\!&\!\left[s
    +\frac{2(s\cos\theta-s)(-n-n\cos\theta)
      -(n^2-n^2\cos^2\theta)}{\sin^2\theta}
  \right]\psi_{0,n}^{s}
  \nonumber\\[0.3\baselineskip]
  \!&\!=\!&\!\left[(2n+1)s-n^2\right] \psi_{0,n}^{s}
  \nonumber\\[0.3\baselineskip]
  \!&\!=\!&\!\left[(2n+1)s_0+n(n+1)\right] \psi_{0,n}^{s},
\end{eqnarray} 
%
which completes the proof.

Note that since $\bs{S}$ commutes with $\bs{L}$, Haldane's coherent
states remain coherent as we elevate them into higher Landau levels.
In particular, the state
\begin{equation}
  \label{eq:Coherent_n}
  \psi_{(\alpha,\beta),n}^{s}(u,v) = (S^-)^n\,(\bar\alpha u + \bar\beta v)^{2s}
\end{equation}
in the $(n+1)$-th Landau level still satisfies \eqref{eq:CoherentDef}.

\section{Pseudopotentials} 

Haldane~\cite{haldane83prl605} also introduced two-particle 
coherent lowest Landau level states defined by
\renewcommand{\strut}{\rule[-.6\baselineskip]{0pt}{2\baselineskip}}
\begin{equation}
  \label{eq:twoCoherentDef}
  \{\bs{\Omega}(\alpha,\beta)\,(\bs{L}_1+\bs{L}_2)\}\,
  \psi_{(\alpha,\beta),0}^{s,j}[u,v]
  =j\,\psi_{(\alpha,\beta),0}^{s,j}[u,v],
\end{equation}
where $[u,v]:=(u_1,u_2,v_1,v_2)$ and $j$ is the total
angular momentum:
\begin{equation}
  \label{eq:twoCoherentJtot}
  (\bs{L}_1+\bs{L}_2)^2\,
  \psi_{(\alpha,\beta),0}^{s,j}[u,v]
  =j(j+1)\,\psi_{(\alpha,\beta),0}^{s,j}[u,v].
\end{equation}
The solution of \eqref{eq:twoCoherentDef} is given by
\begin{equation}
  \label{eq:twoCoherent}
  \psi_{(\alpha,\beta),0}^{s,j}[u,v] = 
  (u_1 v_2 - u_2 v_1)^{2s-j}\,
  \prod_{i=1,2}(\bar\alpha  u_i + \bar\beta  v_i)^{j}.
\end{equation}
It describes two particles with relative momentum ${2s-j}$ precessing
about their common center of mass at $\bs{\Omega}(\alpha,\beta)$.  It
is straightforward to elevate this state into the $(n+1)$-th Landau
level:
\begin{equation}
  \label{eq:twoCoherent_n}
  \psi_{(\alpha,\beta),n}^{s,j}[u,v] 
  = \prod_{i=1,2}(S^-)^n\,\psi_{(\alpha,\beta),0}^{s,j}[u,v]
\end{equation}
Note that \eqref{eq:twoCoherent_n} still satisfies
\eqref{eq:twoCoherentDef} and \eqref{eq:twoCoherentJtot}.

Since $0\le j\le 2s$, the relative momentum quantum number $l=2s-j$
has to be a non-negative integer.  
For bosons or fermions, $l$ has to be even or odd, respectively.  This
implies that the projection $\Pi_n$ onto the $(n+1)$-th Landau level
of any translationally invariant (\ie rotationally invariant on the
sphere) operator $V(\bs{\Omega}_1\cdot\bs{\Omega}_2)$, such as two
particle interaction potentials, can be expanded as
\begin{equation}
  \label{eq:LLpro}
  \Pi_n V(\bs{\Omega}_1\cdot\bs{\Omega}_2) \Pi_n
  =\sum_{l}^{2s} V_l^n\, P_{2s-l}(\bs{L}_1+\bs{L}_2),
\end{equation}
where the sum over $l$ is restricted to even (odd) integer for bosons
(fermions), $P_j(\bs{L})$ 
is the projection operator on states with total momentum $\bs{L}^2=j(j+1)$,
and the $V_l^n$ are pseudopotential coefficients.

The pseudopotential $V_l^n$ denotes the potential energy cost of
$V(\bs{\Omega}_1\cdot\bs{\Omega}_2)$ for two particles with relative
angular momentum $l$ in the $(n+1)$-th Landau level.  We can use
the coherent states \eqref{eq:twoCoherent_n} to evaluate them.  As the
result will not depend on the center of rotation, we can take
$(\alpha,\beta)=(1,0)$, \ie work with the coherent states
\begin{equation}
  \label{eq:twoCoherent_n1}
  \psi_{(1,0),n}^{s,j}[u,v] =
  (S_1^-)^n(S_2^-)^n
  (u_1 v_2 - u_2 v_1)^{2s-j} u_1^j u_2^j. 
\end{equation}
This yields  
\begin{equation}
  \label{eq:Vln}
  V_{2s-j}^n
  =\frac{\bra{\psi_{\!(1,0),n}^{s,j}}V(\bs{\Omega}_1\cdot\bs{\Omega}_2)
  \ket{\psi_{\!(1,0),n}^{s,j}}}
  {\Bigbraket{\psi_{\!(1,0),n}^{s,j}}{\psi_{\!(1,0),n}^{s,j}}}
\end{equation}
for the pseudopotentials.  Since the chord distance between
two points on the unit sphere is given by  
\begin{equation}
  \label{eq:chord}
  \abs{\bs{\Omega}_1-\bs{\Omega}_2}=2\abs{u_1 v_2 - u_2 v_1},
\end{equation}
a $1/r$ or Coulomb interaction on the sphere is given by
\begin{equation}
  \label{eq:VCoulomb}
  V(\bs{\Omega}_1\cdot\bs{\Omega}_2)
  =\frac{1}{2\abs{u_1 v_2 - u_2 v_1}}.
\end{equation}
Fano, Ortolani, and Colombo~\cite{fano-86prb2670} evaluated the
pseudopotential coefficients for Coulomb interactions in the lowest
Landau level by explicit integration, and found
\begin{equation}
  \label{eq:VlCoulomb}
  V_{l}^0=\frac{
    \begin{pmatrix} 2l\\l\end{pmatrix}
    \begin{pmatrix} 8s+2-2l\\4s+1-l\end{pmatrix}}{
    \begin{pmatrix} 4s+2\\2s+1\end{pmatrix}^2}.
\end{equation}


The potential interaction Hamiltonian acting on many particle states
expanded in a basis of $L^z$ eigenstates \eqref{eq:LLLbasis} or
\eqref{eq:LLbasis} is given by
\begin{eqnarray}
  \label{eq:qhsVham2s}
  \hspace{-25pt}H_{\s\text{int}}^{(n)}
  \hspace{-2pt}&=&\hspace{-2pt} 
  \sum_{m_1=-s}^s\, \sum_{m_2=-s}^s\, \sum_{m_3=-s}^s\, \sum_{m_4=-s}^s
  \ a_{m_1,n}^\dagger\, a_{m_2,n}^\dagger\, a_{m_3,n}\, a_{m_4,n}\
  \nonumber\\[0.2\baselineskip]
  &&\hspace{-8pt}
  \cdot\,\delta_{m_1+m_2,m_3+m_4}\sum_{l=0}^{2s}
  \braket{s,m_1;s,m_2}{2s-l,m_1+m_2} V_l^n
  \nonumber\\[0.2\baselineskip]
  &&\hspace{61pt}
  \braket{2s-l,m_3+m_4}{s,m_3;s,m_4}\!,
  \hspace{8pt}
\end{eqnarray}
where $a_{m,n}$ annihilates a boson or fermion in the properly normalized
single particle state
\begin{equation}
  \label{eq:qhsnormLLLbasis}
  \psi_{m,n}^{s}(u,v) = C_{m,n}\, (S^-)^n\,
  u^{s+m}v^{s-m}
\end{equation}
with
\begin{equation}
  \label{eq:Cmn}
  C_{m,n}=\sqrt{\frac{(2s-n)!}{(2s)!\, n!}} 
  \sqrt{\frac{(2s+1)!}{4\pi\,(s+m)!\, (s-m)!}}.
\end{equation}
In \eqref{eq:qhsVham2s}, we take two particles with $L_z$ eigenvalues
$m_3$ and $m_4$, use the Clebsch--Gordan coefficients~\cite{Baym69}
$\braket{2s-l,m_3+m_4}{s,m_3;s,m_4}$ to change the basis into one where
$m_3+m_4$ and the total two particle momentum $2s-l$ are replacing the
quantum numbers $m_3$ and $m_4$, multiply each amplitude by $V_l^n$, and
convert the two particles states back into a basis of $L_z$
eigenvalues $m_1$ and $m_2$.

Note that since this basis transformation commutes with $\bs{S_i}$ for
all $i$, \eqref{eq:qhsVham2s} depends on the Landau level index $n$
only through the pseudopotentials.  This means that if we write out
the potential interaction term \eqref{eq:qhsVham2s} in a higher Landau
level, the matrix we obtain is exactly as in the lowest Landau level
for the same value of $s$, except that we have to use the
pseudopotential $V_l^n$ for the $(n+1)$-th Landau level instead of
$V_l^0$.  Note further that the normalization $C_{m,n}$ for the basis
states factorizes into a term which depends only on 
$n$ and a term which depends only on the $L^z$ eigenvalue $m$.  This
follows again from the commutativity of $\bs{S}$ and $\bs{L}$.  It is
hence sufficient to write out the wave function of a quantized Hall
state in the lowest Landau level using the basis states
\eqref{eq:qhsnormLLLbasis} for $n=0$, and use the Hamiltonian matrix
\eqref{eq:qhsVham2s} with the $(n+1)$-th Landau level pseudopotentials
$V_l^n$ to evaluate the interaction energy this state would have if we
were to elevate it into the $(n+1)$-th level with $\prod_i (S_i^-)^n$.
In other words, the only difference between an exact diagonalization
study in a higher Landau as compared to the lowest Landau level is
that we have to use $V_l^n$ instead of $V_l^0$.


The generalization of the pseudopotentials for three- and more
particle interactions~\cite{simon-07prb195306} to higher Landau levels
proceeds without incident.

\section{Filled Landau levels}
The wave function for a filled $(n+1)$-th Landau level for 
$N=2s+1$ particles with $s_0=s-n$ is given by
\begin{equation}
  \label{eq:psiLLn}
  \psi_{n}^{s}[u,v,\bar u,\bar v]
  =\prod_{i=1}^N(S_i^-)^n\,\prod_{i<j}^N(u_iv_j-u_jv_i).
\end{equation}
Except for $n=0$, this does not reduce to any particularly simple form
when we write out all the terms .

We have found, however, a convenient way to write the wave function for $M$
filled Landau levels with index $n=0,\ldots,M-1$ (\ie from the first
to the $M$-th Landau level).  We assume a total of $LM$ particles
labeled by two integers $l=1,...,L$ and $m=1,...,M$, with spinor
coordinates $(u_{lm},v_{lm},\bar u_{lm},\bar v_{lm})$.  The $LM$
particle wave function for a sphere with $2s_0=L-M>0$ flux quanta is
then given by
\begin{eqnarray}
  \label{eq:psiMLL}
  \psi^{s_0}[u,v,\bar u,\bar v]=
  \mathcal{A}\Biggl\{\, 
  \prod_{m=1}^M \prod_{l<l'}^L &\!\!(u_{lm}v_{l'm}-u_{l'm}v_{lm})\!&\!\!\Biggr.
  \nonumber\\[0.3\baselineskip]
  \cdot\prod_{l=1}^L \prod_{m<m'}^M
  \!&\!\!(\bar u_{lm}\bar v_{lm'}-\bar u_{lm'}\bar v_{lm})\!&\!\!
  \Biggl\}\Biggr. ,
\end{eqnarray}
where $\mathcal{A}$ denotes antisymmetrization.  To verify
\eqref{eq:psiMLL}, multiply the wave functions \eqref{eq:psiLLn}
for each Landau level $n$ with
\renewcommand{\strut}{\rule[-.6\baselineskip]{0pt}{2.5\baselineskip}}
\begin{equation*}
  \prod_{i=1}^N\, (u_i\bar u_i+v_i\bar v_i)^{M-1-n}, 
\end{equation*}
which is equal to 1 and commutes with both $\bs{S}_i$ and $\bs{L}_i$
for all $i$, and then antisymmetrize over all the single particle
states in the Landau levels with index $n=0,\ldots,M-1$.

The formulation \eqref{eq:psiMLL} may be useful in the construction of
composite fermion states\cite{Jain07} for hierarchical filling
fractions, and in particular as a starting point for
obtaining such states from several filled Landau levels through a
process of adiabatic localization of magnetic flux onto the particles
\cite{greiter-90mplb1063}.

\section{Conclusion}
We have developed a formalism to describe the Hilbert space of charged
particles on a sphere subject to a magnetic monopole field, using two
mutually commuting SU(2) algebras for cyclotron and guiding center
momenta.  As the previously developed formalism for the lowest Landau
level has been highly important for numerical studies of fractionally
quantized Hall states, we expect our generalization to higher Landau
levels to be of similar significance.

\appendix
\section{Spherical Coordinates}
The formalism 
requires vector analysis in spherical coordinates.  In this appendix,
we will briefly review the conventions.

Vectors and vector fields are given by
\begin{gather}
  \label{eq:app-spr}
  \bs{r}=r \bs{e}_r, \\
  \bs{v}(\bs{r})= v_r\bs{e}_r + v_\theta\bs{e}_\theta + v_\varphi\bs{e}_\varphi,
\end{gather}
with
\begin{equation}
  \label{eq:app-spe}
  \bs{e}_r = \left(
    \begin{array}{c}
      \cos\varphi\,\sin\theta \\
      \sin\varphi\,\sin\theta \\
      \cos\theta
    \end{array}\right),\  
  \bs{e}_\theta = \left(
    \begin{array}{c}
      \cos\varphi\,\cos\theta \\
      \sin\varphi\,\cos\theta \\
      -\sin\theta
    \end{array}\right),\
  \bs{e}_\varphi = \left(
    \begin{array}{c}
      -\sin\varphi\\
      \cos\varphi \\
      0
    \end{array}\right),
\end{equation}
where $\varphi\in[0,2\pi[$ and $\theta\in[0,\pi]$.  This implies
\begin{equation}
  \label{eq:app-sprhs}
  \bs{e}_r\times\bs{e}_\theta=\bs{e}_\varphi,\quad
  \bs{e}_\theta\times\bs{e}_\varphi=\bs{e}_r,\quad
  \bs{e}_\varphi\times\bs{e}_r=\bs{e}_\theta,
\end{equation}
and
\begin{align}
  \parder{\bs{e}_r}{\theta}&=\bs{e}_\theta,&\!\!
  \parder{\bs{e}_\theta}{\theta}&=-\bs{e}_r,&\!\!
  \parder{\bs{e}_\varphi}{\theta}&=0,\nonumber\\[0.5\baselineskip]
  \parder{\bs{e}_r}{\varphi}&=\sin\theta\,\bs{e}_\varphi,&\!\!
  \parder{\bs{e}_\theta}{\varphi}&=\cos\theta\,\bs{e}_\varphi,&\!\!
  \parder{\bs{e}_\varphi}{\varphi}&=-\sin\theta\,\bs{e}_r-\cos\theta\,\bs{e}_\theta.
  \label{eq:app-spderivatives}
\end{align}
With the nabla operator
\begin{equation}
  \label{eq:app-spnabla}
   \nabla\;=\;\bs{e}_r\parder{}{r}
   +\bs{e}_\theta\frac{1}{r}\parder{}{\theta}
   +\bs{e}_\varphi\frac{1}{r\sin\theta}\parder{}{\varphi}
\end{equation}
we obtain
\begin{eqnarray}
  \label{eq:app-spdiv}
     \nabla \bs{v}\!&\!=\!&\!\frac{1}{r^2}\parder{(r^2v_r)}{r}
     +\frac{1}{r\sin\theta}\parder{(\sin\theta v_\theta)}{\theta}
     +\frac{1}{r\sin\theta}\parder{v_\varphi}{\varphi},
     \\[0.5\baselineskip]\label{eq:app-sprot}
     \nabla\times\bs{v}\!&\!=\!&\!
     \bs{e}_r\frac{1}{r\sin\theta}\left(\parder{(\sin\theta v_\varphi)}{\theta}
       -\parder{v_\theta}{\varphi}\right)\nonumber\\*[0.5\baselineskip]
     \!&\!+\!&\!
     \bs{e}_\theta\left(\frac{1}{r\sin\theta}\parder{v_r}{\varphi}
       -\frac{1}{r}\parder{(rv_\varphi)}{r}\right)\nonumber\\*[0.5\baselineskip]
     \!&\!+\!&\!
     \bs{e}_\varphi\left(\frac{1}{r}\parder{(rv_\theta)}{r}
       -\frac{1}{r}\parder{v_r}{\theta}\right),
     \\[0.5\baselineskip]\label{eq:app-splaplace}
     \nabla^2\!&\!=\!&\!
     \frac{1}{r^2}\parder{}{r}\left(r^2\parder{}{r}\right)
     +\frac{1}{r^2\sin\theta}\parder{}{\theta}
     \left(\sin\theta\parder{}{\theta}\right)\nonumber\\*[0.5\baselineskip]
     \!&\!+\!&\!
     \frac{1}{r^2\sin^2\theta}\parder{^2}{\varphi^2}.\quad
\end{eqnarray}\\[-.5\baselineskip]

Comparing \eqref{eq:lambda^2} with \eqref{eq:app-splaplace}, we see that
\begin{equation}
  \label{eq:qhcomparison}
  \bs{\Lambda}^2\bigl\vert_{s_0=0}\bigr.=\nabla^2\bigl\vert_{r\equiv 1}\bigr.,
\end{equation}
as expected.
 

\end{document}